\documentclass[]{aastex631}

\usepackage{amssymb}
\usepackage{amsthm}
\usepackage{upgreek}
\usepackage{physics}
\usepackage{textcomp, gensymb}
\usepackage{float}
\usepackage{svg}
\usepackage[normalem]{ulem}
\useunder{\uline}{\ul}{}
\usepackage{enumerate}
\newtheorem{innercustomgeneric}{\customgenericname}
\providecommand{\customgenericname}{}
\newcommand{\newcustomtheorem}[2]{%
  \newenvironment{#1}[1]
  {%
   \renewcommand\customgenericname{#2}%
   \renewcommand\theinnercustomgeneric{##1}%
   \innercustomgeneric
  }
  {\endinnercustomgeneric}
}

\newcustomtheorem{customthm}{Theorem}
\newcustomtheorem{customlemma}{Lemma}
\newcustomtheorem{customcorollary}{Corollary}

\begin{document}

\title{On the Distribution of Probe Traffic Volume Estimated without Trajectory Reconstruction}

\correspondingauthor{Gulshan Noorsumar}

\author[0000-0002-2165-747X]{Kentaro Iio}
\affiliation{
Imazucho Otomo \\ Takashima, Shiga 520-1635, Japan}

\author[0000-0002-6718-4508]{Gulshan Noorsumar}
\affiliation{Department of Engineering Sciences, University of Agder \\ 
Jon Lilletuns vei 9 \\ 4879 Grimstad, Norway}
\email{$gulshan.noorsumar@uia.no$}

\author[0000-0002-7434-6886]{Dominique Lord}
\affiliation{Zachry Department of Civil and Environmental Engineering, Texas A\&M University \\
3127 TAMU \\
College Station, Texas 77843-3127, United States}

\author[0000-0003-2404-5409]{Yunlong Zhang}
\affiliation{Zachry Department of Civil and Environmental Engineering, Texas A\&M University \\
3127 TAMU \\
College Station, Texas 77843-3127, United States}




\begin{abstract}
In recent years, passively recorded probe traffic volumes have increasingly been used to estimate traffic volumes.
However, it is not always possible to count probe traffic volume in a spatial dataset when probe trajectories cannot be fully reconstructed from raw probe point location data due to sparse recording intervals, lack of pseudonyms or timestamps. As a result, the application of such probe point location data has been limited in traffic volume estimation. To relax these constraints, we present the exact distribution of the estimated probe traffic volume in a road segment based on probe point location data without trajectory reconstruction. The distribution of the estimated probe traffic volume can exhibit multimodality, without necessarily being line-symmetric with respect to the true probe traffic volume. As more probes are present, the distribution approaches a normal distribution. The conformity of the distribution was visualised through numerical simulations. Sometimes, there exists a local optimal cordon length that maximises estimation precision. The theoretical variance of estimated probe traffic volume can address heteroscedasticity in the modelling of traffic volume estimates.
\end{abstract}


\keywords{Probe Data, Point Data, Traffic Volume, AADT, Telematics, Privacy Protection, Trajectory Reconstruction}



\section{Introduction} \label{sec:intro}

Traffic volume is a fundamental element of transportation engineering \citep{GREENSHIELDS-1934}, urban planning, real estate valuation, air pollution models \citep{LURIAetal-1990, OKAMOTOetal-1990}, wildlife protection \citep{SEILERandHELLDIN-2006}, and marketing \citep{ALEXANDERetal-2005}. Traffic counts are typically performed at fixed locations using equipment such as pneumatic tubes, loop coils, radars, ultrasonic sensors, video cameras, and light detection and ranging (LiDAR) systems \citep{ZHAOetal-2019}. While conventional traffic counts are believed to have acceptable precision, traffic counts at fixed locations are constrained in space, time, and budget. For this reason, average annual daily traffic (AADT), which is one of the basic traffic metrics in traffic engineering, is often estimated based on 24- or 48-h traffic counts with temporal adjustments \citep{JESSBERGERetal-2016, KRILE-2016, RITCHIE-1986}. Nevertheless, this scalability constraint still places transportation professionals on a leash. For example, researchers have pointed out a lack of reliable traffic volume data in substantive road safety analyses \citep{CHENetal-2019, ELBASYOUNY-2010, MITRAandWASHINGTON-2012, ZAREIandHELLINGA-2022}. 

To maximise the value of limited numbers of traffic counts, extensive research efforts have been devoted to developing traffic volume estimation methods focused on calibration and its accuracy. Such approaches include travel demand modelling \citep{ZHONGandHANSON-2009}, spatial kriging \citep{SELBYandKOCKELMAN-2013}, support vector machines \citep{SUNandDAS-2015}, linear and logistic regressions \citep{APRONTIetal-2016}, geographically weighted regression \citep{PULUGURTHAandMATHEW-2021}, locally weighted power curves \citep{CHANGandCHEON-2019}, and clustering \citep{SFYRIDISandAGNOLUCCI-2020}.

\subsection{Probe Data in Traffic Volume Estimation} \label{subsec:probedataintrafficvolumeestimation}

With the advancements in information technology, expectations for traffic volume availability have increased. In the United States, for example, the Highway Safety Improvement Program (HSIP) asks state departments of transportation to prepare traffic volume data even on low-volume roads \citep{FHWA-2016}. As mobile devices compatible with global navigation satellite systems (GNSSs) have spread throughout our daily lives, opportunities to estimate traffic volumes based on passively collected location data have gained industry attention \citep{CACERESetal-2008, HARRISONetal-2020}. Road agencies have started exploring the feasibility of using probe data to estimate traffic volumes \citep{CODJOEetal-2020, FISHetal-2021, KRILEandSLONE-2021, MACFARLANE-2020, ZHANGetal-2019} because probe traffic volumes and non-probe traffic volumes tend to be positively correlated. In proprietary products providing AADT estimations, reports have found negative correlations between true traffic volumes and estimation accuracy as measured by percentage errors \citep{BARRIOSandCASBURN-2019, ROLL-2019, SCHEWELetal-2021, TSAPAKISetal-2020, TSAPAKISetal-2021a, TURNERetal-2020, YANGetal-2020}.

Machine learning methods have become popular calibration tools for traffic volume estimation based on probe location data. For instance, \cite{MENGetal-2017} and \cite{ZHANetal-2017} applied spatio-temporal semi-supervised learning and an unsupervised graphical model, respectively, to taxi trajectories in Chinese cities to estimate citywide traffic volumes. With a Maryland probe dataset, \cite{SEKULAetal-2018}, for example, showed that neural networks could significantly improve estimation accuracy. In Kentucky, \cite{ZHANGandCHEN-2020} used annual average daily probes (AADP) and betweenness centrality to estimate AADTs across the state. Using random forest, they found that an AADP of 53 was the lower threshold for having a mean absolute percentage error (MAPE) of less than 20\% to 25\%. \cite{SCHEWELetal-2021} reported that gradient boosting excelled in calibrating probe location data for traffic volume estimation.

\subsection{Types of Probe Data} \label{subsec:typefofprobedata}

\begin{figure}[ht]
    \gridline{\fig{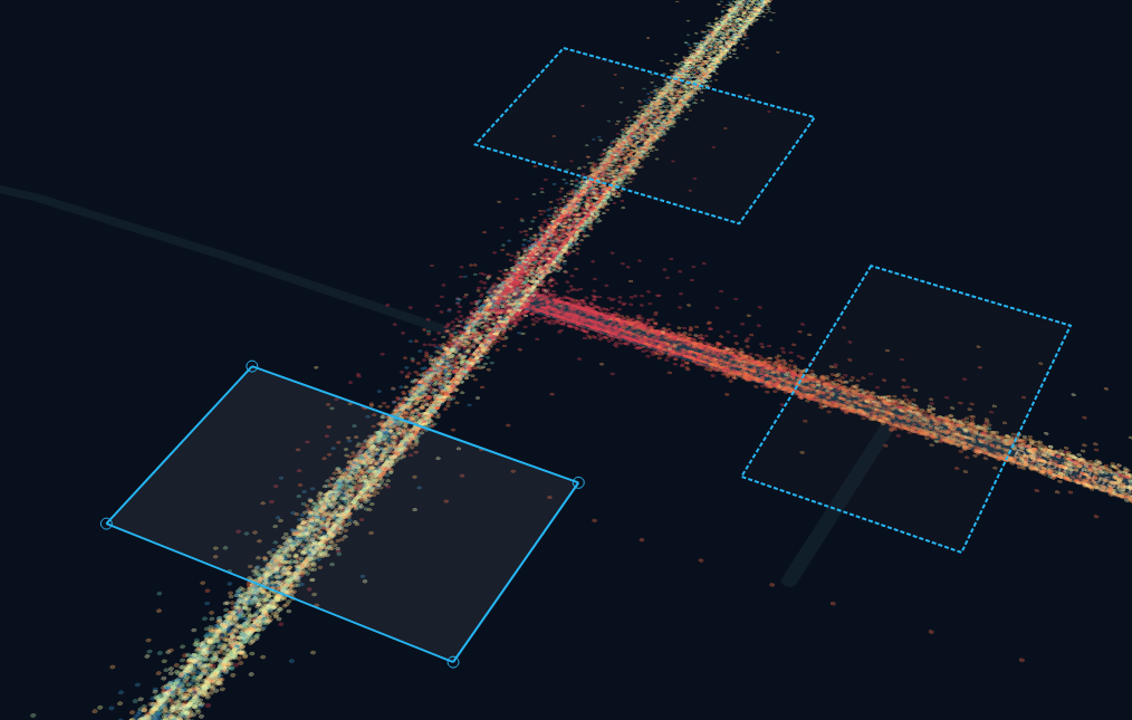}{0.43\textwidth}{(a) Virtual cordons over probe point data \label{fig:figure1a}}
              \fig{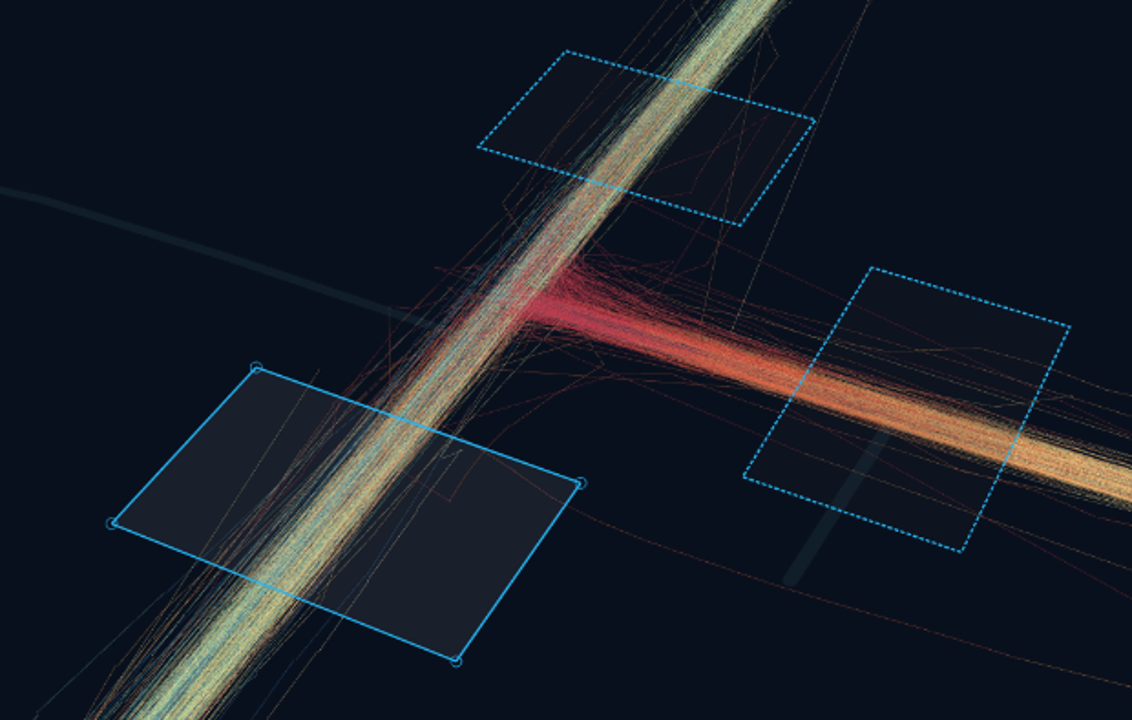}{0.43\textwidth}{(b) Virtual cordons over probe line data \label{fig:figure1b}}}
    \caption{Illustrated virtual cordons over probe point data and line data (reconstructed trajectories).
    \label{fig:figure1}}
\end{figure}

Figure \ref{fig:figure1} illustrates different types of probe data: point data (Figure \ref{fig:figure1}a) and line data (Figure \ref{fig:figure1}b). Point data refer to data that contain information to identify a point location (e.g., geographic coordinates) on a surface, such as the Earth’s ellipsoid. Location data are usually first recorded and stored as point data. In contrast, line data, also called trajectories, paths, or routes, consist of a series of point data of an entity connected chronologically \citep{MARKOVICetal-2019}. Conventional traffic counts require information on passing objects over a cross-section at a fixed location. With probe data, one can count the number of probes passing through a specific location based on trajectories reconstructed from point data (e.g., GPS Exchange Format (GPX)) when the point data meet all of the following conditions:

\begin{itemize}
    \item Each probe has a pseudonym (e.g., device identifier).
    \item Each point data has a timestamp in the ordinal scale or a higher level of measurement.
    \item The recording interval is small enough to determine a route.
\end{itemize}

In other words, data that meet these conditions have less anonymity because one can track each probe’s locations and time simultaneously \citep{DEMONTJOYEetal-2013}. In fact, all of the aforementioned studies used line data of probes to estimate traffic volumes. However, some point data, such as sparsely recorded probe data \citep{SUNetal-2013}, are unsuitable for the precise reconstruction of line data. In addition, agencies might not be able to obtain detailed line data in which they can identify a probe’s geographic coordinates and timestamps at once, depending on privacy regulations and data providers’ policies.

If the number of passing probes can be estimated based on sparse, nonchronological probe point data without pseudonyms, one will be able to use the estimated probe traffic volumes to further estimate traffic volumes. To relax these probe data availability constraints, this paper presents a method for estimating passing probe traffic volumes using point location data collected from the probes without route reconstruction. 
In the following sections, we describe the exact distribution of the unbiased estimator that allows one to assess the estimation precision. We derive analytical relationships between probe traffic variables and estimated probe counts with example calculations. Numerical simulations visualise the conformity of the distribution. Finally, we discuss the characteristics, limitations, applications, and opportunities of the model. It should be noted that we will hardly tap into detailed calibration methods against known traffic volumes because the calibration methods are not essentially unique to this paper.


\section{Theory} \label{sec:theory}
This section describes the problem, provides our findings with proofs, and offers illustrative examples. The examples are provided solely to aid the reader's understanding and are neither the basis for the conclusions of this paper nor a limitation on the situations to which the proposed equations can be applied. We adhere to the International System of Units throughout the paper unless stated otherwise.

\subsection{Problem Statement} \label{subsec:problemstatement}

We define a “probe” as a device that records its position as point data in the Earth's spatial reference system (e.g., geographic coordinates). For instance, a smartphone or connected vehicle can serve as a probe. We want to estimate the number of probes that traversed a road segment during an observation period. Let $m \in \mathbf{Z}^{nonneg}$ and $\hat{m}$ denote the true number of probes passing through a unit segment during an observation period and its estimator, respectively. We present the distribution of $\hat{m}$ under the following conditions.

Assume that each probe traverses the Earth's surface at a space-mean speed \citep{TURNERetal-1998} of $S \in \mathbf{R}^{+}$ m/s, where $S$ is an independent and identically distributed (i.i.d.) continuous random variable\footnote{Because the order of recorded point location data is exchangeable after they are recorded, $S$ can be considered a random variable emerging from the underlying i.i.d. $g(s)$ \citep{DEFINETTI-1930}. Please note that $s$ is not necessarily the same as free-flow speed or target speed.}. We denote the realised value of $S$ as $s \in \mathbf{R}^{+}$. Let $g(s) \in \mathbf{R}^{nonneg} \mid 0 \leq g(s) \mid \displaystyle \int_{0}^{\infty} g(s)\mathrm{d}s = 1$ be the probability density function (PDF) of the probe speed population, a hypothetical infinite group of $s$, within a cordon.
The possibility that multiple probes may be carried by one vehicle at the same space-mean speed $s$ is accounted for in $g(s)$. All probes share the same data recording interval $t \in \mathbf{R}^{+}$ s. Because all speeds are considered in $g(s)$, virtual uniform motion is assumed in the modelling process. Note that assuming virtual uniform motion is different from assuming all probes actually traverse the segment with a uniform motion; rather, it means that any changes in space-mean speed among all probes over the segment are reflected in $g(s)$. In a uniform motion, each probe records its position and space-mean speed as point data (i.e., “\textit{footprints}”) at an interval of $t$ s. Probe identifiers $i$ or detailed timestamps are not necessarily recorded, but data points have at least nominal information to identify a recorded time range of interest (e.g., a label of “July 2023”). We assume no errors or failures in the positioning or recording in the formulation.

\begin{figure}[ht]
    \centering
    \includegraphics[width=0.88\textwidth]{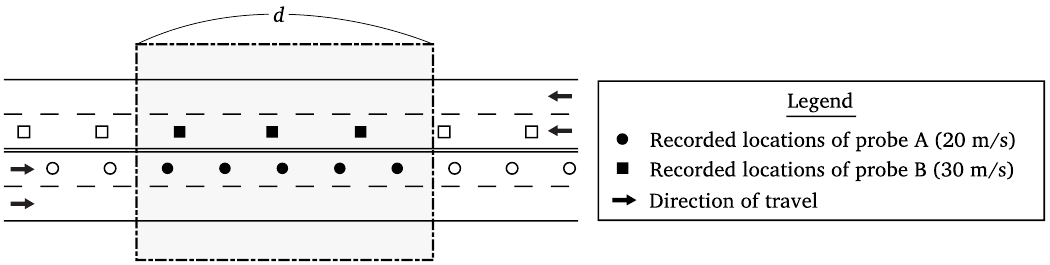}
    \caption{An illustrated example of a virtual cordon over point data ($m = 2$).}
    \label{fig:figure2}
\end{figure}

An analyst draws a $d$-m virtual cordon ($ d \in \mathbf{R}^{+}$) over the data measured along the road segment of interest. This spatial data cropping results in each probe recording its first location in the virtual cordon at a uniformly distributed random time within $t$ s after the probe enters the cordon. The analyst may extract data within the time range of interest as needed. The virtual cordon will contain $n \in \mathbf{Z}^{nonneg}$ data points at a speed of $s_a$ where $a \in \mathbf{Z}^{nonneg} \mid a \leq n$ is a record identifier. Figure \ref{fig:figure2} shows an example of a virtual cordon capturing eight point location data during an observation period. Although the figure differentiates between the two probes, this work does not assume that analysts have information to identify individual probes.

\subsection{Unbiased Estimator of $m$} \label{subsec:lemma1}

\begin{customlemma}{1}\label{lma:lemma1}\normalfont
    If we define $\hat{m}$ as
    
    \begin{equation} \label{eq:draso}
        \hat{m} = \frac{t}{d} \sum_{a=1}^{n}s_{a}, \forall m, d, t, n, s
    \end{equation}
    
    \noindent
    $\hat{m}$ is an unbiased estimator of the true probe traffic volume $m$ (Equation \ref{eq:kgbnho}).
    
    \begin{equation} \label{eq:kgbnho}
        \mathrm{E}[\hat{m}] = m, \forall m
    \end{equation}
\end{customlemma}

\begin{quote}
    \begin{proof}
        Because uniform motion is virtually assumed, $s_i = s_a$ for any probe and $s_it$ is the distance the $i$th probe traverses in $t$ s. Using $n_i$ as the number of data points within a cordon from the probe, Equation \ref{eq:draso} can be reduced to
    
    \begin{equation} \label{eq:uthnbgn}
        \hat{m} = \displaystyle \frac{t s_i n_i}{d}
    \end{equation}
    
        \noindent
        for the $i$th probe. In Equations \ref{eq:draso} and \ref{eq:uthnbgn}, $n_i$ can be broken down into $n_i = \tilde{n}_i + K_i$ where $\tilde{n}_i \in \{ \tilde{n} \in \mathbf{Z}^{nonneg} \}$ is the minimum number of data points that could be recorded in the virtual cordon. It is calculated with the floor function as   
          
        \begin{equation} \label{eq:kdtfg2}
            \tilde{n}_i = \left\lfloor\frac{d}{s_i t}\right\rfloor
        \end{equation}
        
        \noindent
        Here, $K_i$ is a Bernoulli random variable representing the number of additional data points per probe $K_i \in \{ K \in \mathbf \{0, 1\} \}$ observed in addition to $\tilde{n}_i$ data points. Because uniform motion is assumed and a probe leaves its first record in the cordon at a random time within $t$ s after entering the cordon. Naturally, an additional data point is recorded at the probability equal to the fractional part of $d/(s_it)$. When we define the fractional part as $p_i \in \{ p \in \mathbf{R}^{nonneg} \mid 0 \leq p < 1 \}$,
        
        \begin{equation} \label{eq:yathe2}
            p_i = \frac{d}{s_it} \mod 1
        \end{equation}
    
        Because $K_i$ follows the Bernoulli distribution $Ber(p_i)$, its expected value $\mathrm{E}[K_i]$ is $p_i$. From Equations \ref{eq:uthnbgn}, \ref{eq:kdtfg2}, and \ref{eq:yathe2}, $\mathrm{E}[\hat{m}]$, the expected value of $\hat{m}$, is
        \begin{equation} \label{eq:yatlhh2}
            \mathrm{E}[\hat{m}] = \frac{s_it}{d} \left[ \left\lfloor\frac{d}{s_it}\right\rfloor + \left( \frac{d}{s_it} \mod 1 \right) \right] = 1
        \end{equation}
    
        \noindent
         when $m = 1$. Accordingly, $\mathrm{E}[\hat{m}] = m$ for any $m$. Therefore, $\hat{m}$ is an unbiased estimator of $m$.
    \end{proof}
\end{quote}

\subsubsection{Example 1} \label{subsub:example1}

We assume $d = 100$ and $t = 1$ in Figure \ref{fig:figure2}. The expected number of data points from probe B ($s_i = 30$) within the segment is $100/(30 \cdot 1) \approx 3.333$; therefore, at least three data points are observed (i.e., $\tilde{n} = 3$). Since it is impossible to observe 3.333 data points, one more data point is observed at a probability of approximately 0.333 (i.e., $p_i \approx 0.333$). In Figure \ref{fig:figure2}, $m = 2$, $\mathrm{E}[\hat{m}] = 2$ and $\hat{m} = 1.9$. If the cordon had contained the data points only from probe A, $m = 1$, $\mathrm{E}[\hat{m}] = 1$ and $\hat{m} = 1$. If the cordon had included the data points only from probe B, $m = 1$, $\mathrm{E}[\hat{m}] = 1$ and $\hat{m} = 0.9$.

\subsection{Variance of $\hat{m}$} \label{subsec:lemma2}

\begin{customlemma}{2}\label{lma:lemma2}\normalfont
    When we denote the variance of $\hat{m}$ as $\mathrm{Var}[\hat{m}]$:
    
    \begin{equation} \label{eq:efrus}
         \mathrm{Var}[\hat{m}]=\frac{mt^2}{d^2}\int_0^{\infty} b(s, d, t)g(s)\mathrm{d}s
    \end{equation}
    
    \noindent
    where
    
    \begin{equation} \label{eq:bahos}
        b(s, d, t) = s^2 p(1-p) = s^2 \left( \frac{d}{st} \mod 1 \right) \left[1- \left( \frac{d}{st} \mod 1 \right)\right]
    \end{equation}
\end{customlemma}

\begin{quote}
    \begin{proof}
         The variance of $\hat{m}$ arises from the discreteness of the number of recorded data points, namely, the Bernoulli random variable $K$. From Equation \ref{eq:uthnbgn} and the multiplication rule of probability, $\mathrm{Var}[\hat{m} \mid S = s_i]$ is proportional to the variance of the Bernoulli distribution $p(1-p)$ multiplied by the scaling factor $st/d$ raised to a power of 2. Because $S \sim g(s)$, integrating $s^2 t^2 p(1-p)g(s)/d^2$ over $s$ gives the variance of $\hat{m}$ per probe. Because $S$ is i.i.d.,  $\mathrm{Var}[\hat{m}] \propto m$ due to the additivity of variances.
    \end{proof}
\end{quote}

\subsubsection{Example 2} \label{subsub:example2}

\begin{figure}[ht]
\gridline{\fig{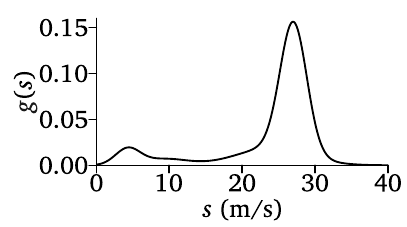}{0.27\textwidth}{(a) $g(s)$ \label{fig:figure3a}}
        \fig{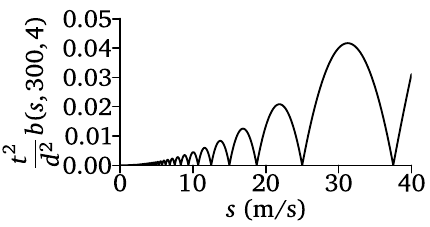}{0.27\textwidth}{(b) $\displaystyle \frac{t^2}{d^2}b(s,300, 4)$ \label{fig:figure3b}}
          \fig{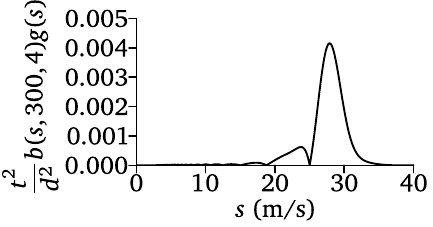}{0.27\textwidth}{(c) $\displaystyle \frac{t^2}{d^2}b(s,300,4)g(s)$ \label{fig:figure3c}}}
\caption{Variance derivation when $d$ = 300, $t$ = 4, and $S \sim g(s)$.
\label{fig:figure3}}
\end{figure}

Hereafter, we use a finite mixture of normal distributions by \cite{PARKetal-2010} as an example of $g(s)$. The speed distribution $g(s)$ had been fitted\footnote{Marginal likelihood was -115,052.3 and Bayes factor was 146.9.} to 24-h speed data collected on Interstate Highway 35 (I-35) in Texas. Capturing 24-h speed variation, the distribution comprises four normal distributions $N(\upmu, \upsigma^2)$ defined by $\upmu = (27.042, 24.000, 9.394, 4.294)$, $\upsigma = (1.831, 4.797, 3.167, 1.686)$, $w = (0.647, 0.223, 0.055, 0.074)$, and $\displaystyle \sum w_j = 1$ where $\upmu \in \mathbf{R}$ is a tuple (i.e., a finite ordered list) of mean speed in m/s, $\upsigma \in \mathbf{R}^{nonneg}$ is a tuple of standard deviation in m/s before truncation, and $w \in \mathbf{R}^{nonneg}\, |\, w \leq 1$ defines the proportions of the normal distributions within the mixture. The distribution was truncated at $s = 0$ and $s = 40$. The resulting $g(s)$ is a mixture of four truncated normal distributions, defined by the following equations (Figure \ref{fig:figure3}a):

\begin{equation} \label{eq:bruxl}
    g(s \mid \upmu, \upsigma, 0, 40) =
    \begin{cases}
    \displaystyle \sum\limits_{j=1}^4
    w_j \uppsi(s \mid \upmu_j, \upsigma_j, 0, 40)
    , & 0 < s \leq 40 \\
    0, & \text{otherwise}
    \end{cases}
\end{equation}

\noindent
where $\upalpha < \upbeta$, $0 < \upsigma$, and

\begin{equation} \label{eq:trunc}
    \uppsi(x \mid \upmu, \upsigma, \upalpha, \upbeta) =
    \frac{\upphi \displaystyle \left( \frac{x-\upmu}{\upsigma} \right)}
    {\upsigma \displaystyle \left[ \Upphi \displaystyle \left( \frac{\upbeta-\upmu}{\upsigma} \right) - \Upphi \displaystyle \left(\frac{\upalpha-\upmu}{\upsigma}\right) \right] } \\
\end{equation}

\begin{equation} \label{eq:iojmh}
    \upphi(x) = \displaystyle \frac{e^{\displaystyle \left( \frac{-x^2}{2} \right) }}{\sqrt{2\uppi}}
\end{equation}

\begin{equation} \label{eq:jkniy}
    \Upphi(x) = \frac{1}{2} \displaystyle \left[1 + \operatorname*{erf}\displaystyle \left(\frac{x}{\sqrt{2}}\right) \right]\\
\end{equation}

Assuming $d = 300$ and $t = 4$, Figure \ref{fig:figure3}b displays $4^2/300^2 \cdot b(s, 300, 4)$, the variance in the estimated probe traffic volume as a function of $s$ (Equation \ref{eq:bahos}). If $S$ were uniformly distributed between 0 and 40 (i.e., $S \sim U(0, 40]$), the area under the function in Figure \ref{fig:figure3}b would have been proportional to the variance of the estimated probe traffic volume (i.e., $\mathrm{Var}[\hat{m} \mid S = s_i]$). Here, we want to weigh $4^2/300^2 \cdot b(s, 300, 4)$ by $g(s)$ because $S \sim g(s)$. This operation results in Figure \ref{fig:figure3}c, where the area under the function, 0.019, is the theoretical variance of $\hat{m}$ from a probe (Equation \ref{eq:efrus}).

\subsection{Shape of $\hat{m}$} \label{subsec:theorem1}

\begin{customthm}{1}\label{theorem:theorem1}\normalfont
    Let $u \in \mathbf{Z}^{nonneg}$ be a nonnegative integer that operationally substitutes $\tilde{n}$. With the previously defined variables and a function, the PDF of $\hat{m}$ is given as $f(\hat{m}; m)$:
    
    \begin{equation} \label{eq:hnmnmi}
        f(\hat{m}; m) = f'^{*m}(\hat{m})
    \end{equation}
    
    \noindent
    where $f'^{*m}(\hat{m})$ denotes $m$-fold self-convolution of $f'(\hat{m})$. The function $f'(\hat{m})$ is defined as
    
    \begin{equation} \label{eq:rojoc}
        f'(\hat{m})=\displaystyle \sum_{u=0}^{\infty}\sum_{k=0}^{1} h(\hat{m}; t, d, u, k) 
    \end{equation}
    
    \noindent
    where
    
    \begin{equation} \label{eq:husan}
        h(\hat{m}; t, d, u, k)  =
        \begin{cases}
        \displaystyle g\left(\frac{d\hat{m}}{t(u+k)}\right)
        \frac{p^{k}(1 - p)^{1-k}d}{t (u+k)}
        , & (u = 0 \land k \neq 0) \lor \left( u \neq 0 \land \displaystyle \frac{u+k}{u+1} < \hat{m} \leq \displaystyle \frac{u+k}{u} \right)\\
        0, & \text{otherwise}
        \end{cases}
    \end{equation}
\end{customthm}

\begin{quote}
    \begin{proof}
        From Equations \ref{eq:kdtfg2} and \ref{eq:yathe2}, $s$ uniquely determines $\tilde{n}$ and $p$ once $d$ and $t$ are determined. In addition, any single $s$ has a mutually exclusive set of $k$ as the outcome of a Bernoulli trial. In Equation \ref{eq:uthnbgn}, $\hat{m}$ is a linear function of $s$ with slope $t (\tilde{n}+k)/d$. Because the probe speed $S$ is i.i.d., the sum of all relative frequencies for possible occurrences of $\tilde{n}$ and $k$ by $\hat{m}$ gives the PDF of $\hat{m}$; therefore, the PDF of $\hat{m}$ contains the joint probability function $g(s)p^{k}(1 - p)^{1-k}$. In Equations \ref{eq:rojoc} and \ref{eq:husan}, $u$ substitutes for $\tilde{n}$. Let $x \in \mathbf{R}^{nonneg}$ be a nonnegative real number and $\updelta$ be an infinitesimal interval. The probability that $\hat{m}$ takes a value in the interval $(x, x + \updelta ]$ is calculated by integrating the PDF of $\hat{m}$ over the interval. From Equation \ref{eq:uthnbgn}, $m = 0$ when $u + k = 0$; otherwise, the interval of $s$ corresponding to $(x, x + \updelta]$ is $(s, s+ \updelta']$ = $\displaystyle \left(dx/\left[t(u+k)\right], dx/\left[t(u+k)\right] + \updelta d/\left[t(u+k)\right] \right]$, where $dx/\left[t(u+k)\right]$ is $s$ as a function of $\hat{m}$  and $d/\left[t(u+k)\right]$ is the reciprocal of the slope of $\hat{m}$ as a function of $s$ (e.g., Figure \ref{fig:figure4}). However, the interval of $s$ must be constant regardless of $\hat{m}$ in the PDF of $\hat{m}$ because $\hat{m}$ results from $S$, but not vice versa. Therefore, the joint probability of $u$ and $k$, in fact, must be multiplied by $d/\left[t(u+k)\right]$, which is the reciprocal of the slope of $s$ as a function of $\hat{m}$. When $S$ is i.i.d., $\hat{m}$ is also i.i.d. (Equation \ref{eq:uthnbgn}). Hence, the PDF of $\hat{m}$ emerges as an $m$-fold self-convolution of the PDF where $m = 1$ (Equation \ref{eq:hnmnmi}).  
    \end{proof}
\end{quote}

\begin{customcorollary}{1}\label{crlry:corollary1}\normalfont
    As $m$ approaches infinity, the shape of $f(\hat{m}; m)$ converges to that of a normal distribution:
    
    \begin{equation} \label{eq:oihnmybn}
        \lim_{{m \to \infty}} f(\hat{m}; m) = N \left(m, \frac{mt^2}{d^2} \int_0^{\infty} b(s,d,t)g(s)\mathrm{d}s\right)
    \end{equation}
\end{customcorollary}

\begin{quote}
    \begin{proof}
        Because $\hat{m}$ is i.i.d., Equation \ref{eq:oihnmybn} is derived from the classical central limit theorem on lemmata \ref{lma:lemma1} and \ref{lma:lemma2}.
    \end{proof}
\end{quote}

\subsubsection{Example 3} \label{subsub:example3}

\begin{figure}[ht]
    \centering
    \includegraphics[width=0.53\textwidth]{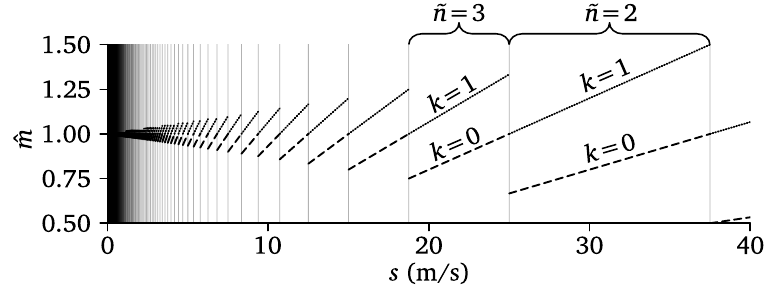}
    \caption{$\hat{m}$ as a function of $s$ and $k$ when $d$ = 300, $t$ = 4, and $m$ = 1.
    \label{fig:figure4}}
\end{figure}

\begin{figure}[ht]
    \centering
    \includegraphics[width=0.53\textwidth]{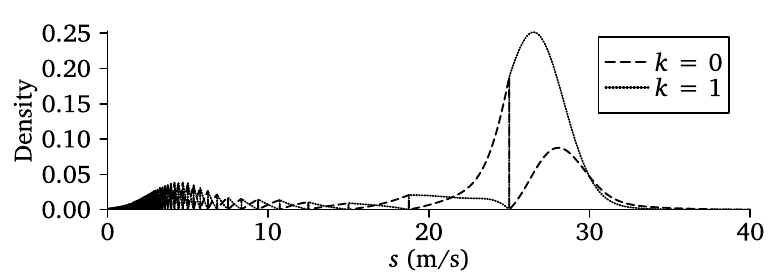}
    \caption{The PMFs of $Ber(p)$ weighted by $g(s)$ as a function of $s$ and $k$ when $d$ = 300 and $t$ = 4.
    \label{fig:figure5}}
\end{figure}

Assuming $d = 300$ and $t = 4$, Figure \ref{fig:figure4} plots Equation \ref{eq:uthnbgn} (i.e., when $m = 1$). The combinations of $\tilde{n}$ and $k$ form an infinite periodic pattern along the $s$-axis because $\tilde{n}$ increases towards infinity as $s$ approaches 0. Because $S \sim g(s)$, we want to take the relative frequency of speed and each $k$ by multiplying the probability mass function (PMF) of $Ber(p)$ by $g(s)$. This operation results in the overall frequency of the combination of $\tilde{n}$ and $k$ by $s$ (Figure \ref{fig:figure5}).

\begin{figure}[ht] 
    \centering
    \includegraphics[width=0.32\textwidth]{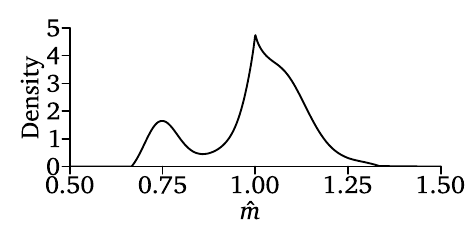}
    \caption{The PDF of $\hat{m}$ when $m = 1, d=300,$ and $t=4$.
    \label{fig:figure6}}
\end{figure}

From Figure \ref{fig:figure4}, it is apparent that the density of $\hat{m}$ can arise from multiple combinations of $\tilde{n}$ and $k$, which have different slopes for $\hat{m}$ with respect to $s$. Therefore, an infinitesimal interval of $\hat{m}$ can have different cardinalities of the frequencies projected from the $s$-axis; thus, we must consider the cardinality of $\hat{m}$. For example, the length of an infinitesimal interval of $\hat{m}$ corresponding to any interval between $s = 25$ and $s = 37.5$ in Figure \ref{fig:figure4} is 50\% longer when $k = 1$ than when $k = 0$. Because we are interested in the PDF of $\hat{m}$, we must normalise the value using the cardinality of $\hat{m}$. This operation can be performed by dividing the relative frequency given the combination of $\tilde{n}$ and $k$ by each slope $t(\tilde{n}+k)/d$ before summation. Equation \ref{eq:rojoc} results in the PDF in Figure \ref{fig:figure6} in this example.

\subsection{Optimal Cordon Length} \label{subsec:theorem2}

Equation \ref{eq:efrus} indicates that $d$ determines $\mathrm{Var}[\hat{m}]$ when $t$ and $g(s)$ are already fixed.  Considering that $d$ is often the only parameter that an analyst can control, the art of estimation error minimisation lies in setting a good cordon length $d$. That said, what length should $d$ be under which conditions? Modelling the relationships between $\hat{m}$ and the other variables gives us a hint on choosing a good cordon length $d$. 

\begin{customcorollary}{2}\label{crlry:corollary2}\normalfont
    Let $\max(d)$ denote the maximum feasible $d$ within a given segment. When $\max(d)$ exists, there can be a cordon length $d$ shorter than $\max(d)$ that minimises the precision of estimating $m$. Such a value of $d$ can be sought by $\displaystyle \operatorname*{argmin}_{0 < d \leq \max(d)} obj(d)$ where $obj(d)$ is an objective function such as the variance-to-mean ratio (VMR)
    
    \begin{equation} \label{eq:ghvry}
    \mathrm{VMR}[\hat{m}] = 
    \frac{\mathrm{Var}[\hat{m}]}{\mathrm{E}[\hat{m}]} = \frac{t^2}{d^2}\int_0^{\infty} b(s,d,t)g(s)\mathrm{d}s
    \end{equation}
    
    \noindent
    or the coefficient of variation (CV)
    
    \begin{equation} \label{eq:hokeg}
    \mathrm{CV}[\hat{m}] = 
    \frac{ \sqrt{\mathrm{Var}[\hat{m}]} }{\mathrm{E}[\hat{m}]} = \frac{t}{d} \sqrt{\frac{1}{m} \int_0^{\infty} b(s,d,t)g(s)\mathrm{d}s}
    \end{equation}
\end{customcorollary}

\begin{quote}
    \begin{proof}
    Assume that Corollary 2 is false.
    When $m = 1$, $t = 4$ and $S \sim g(s)$ defined by Equations \ref{eq:bruxl}-\ref{eq:jkniy}, $\mathrm{CV}[\hat{m}] = 0.310$ when $d = 150$ whereas $\mathrm{CV}[\hat{m}] = 0.230$ when $d = 110$. Because there is a counterexample to the assumption that Corollary 2 is false, Corollary 2 is true.
    \end{proof}
\end{quote}

\subsubsection{Example 4} \label{subsub:example4}

\begin{figure}[H] 
\gridline{\fig{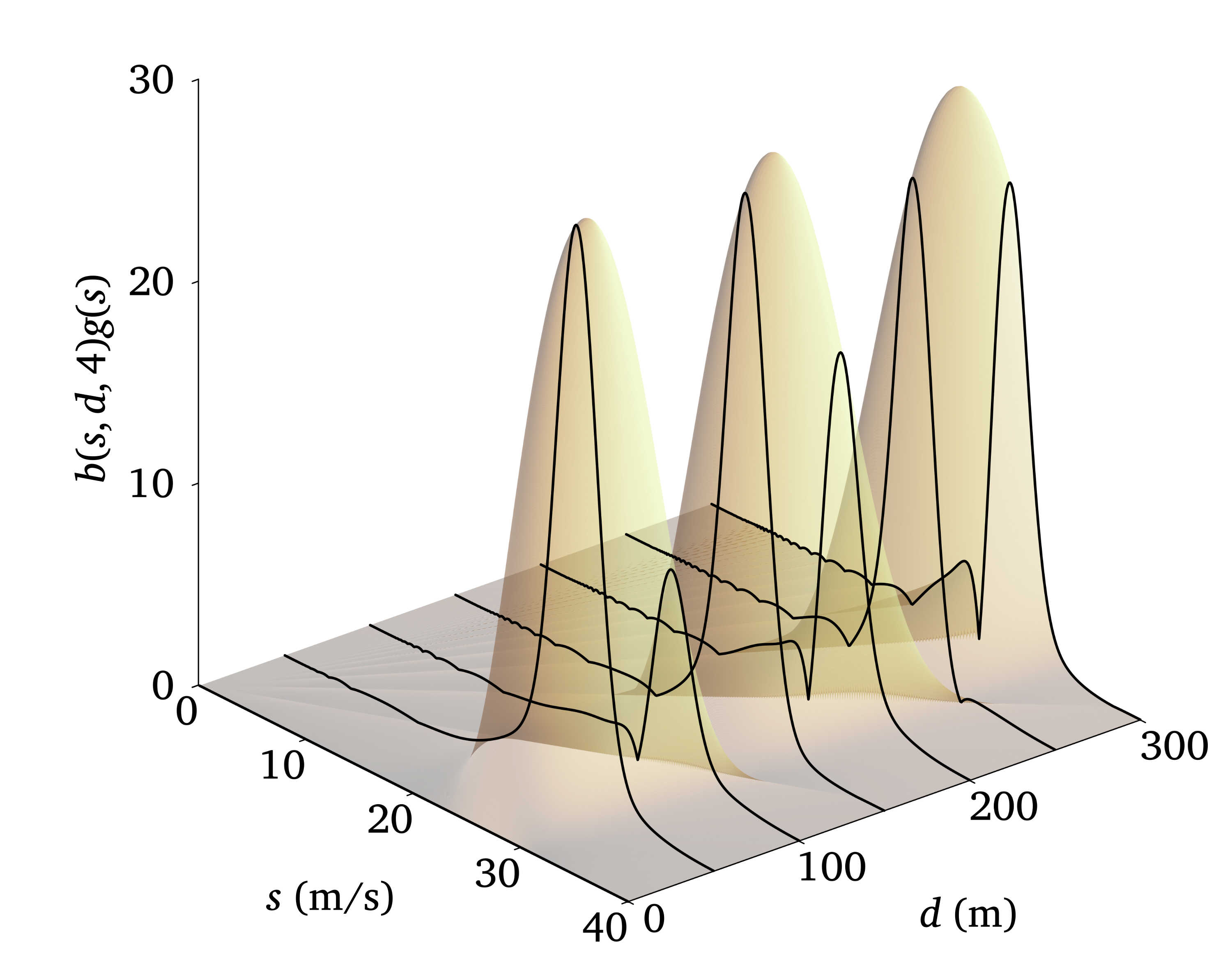}{0.49\textwidth}{(a) $b(s, d, 4)g(s)$ \label{fig:figure7a}}
          \fig{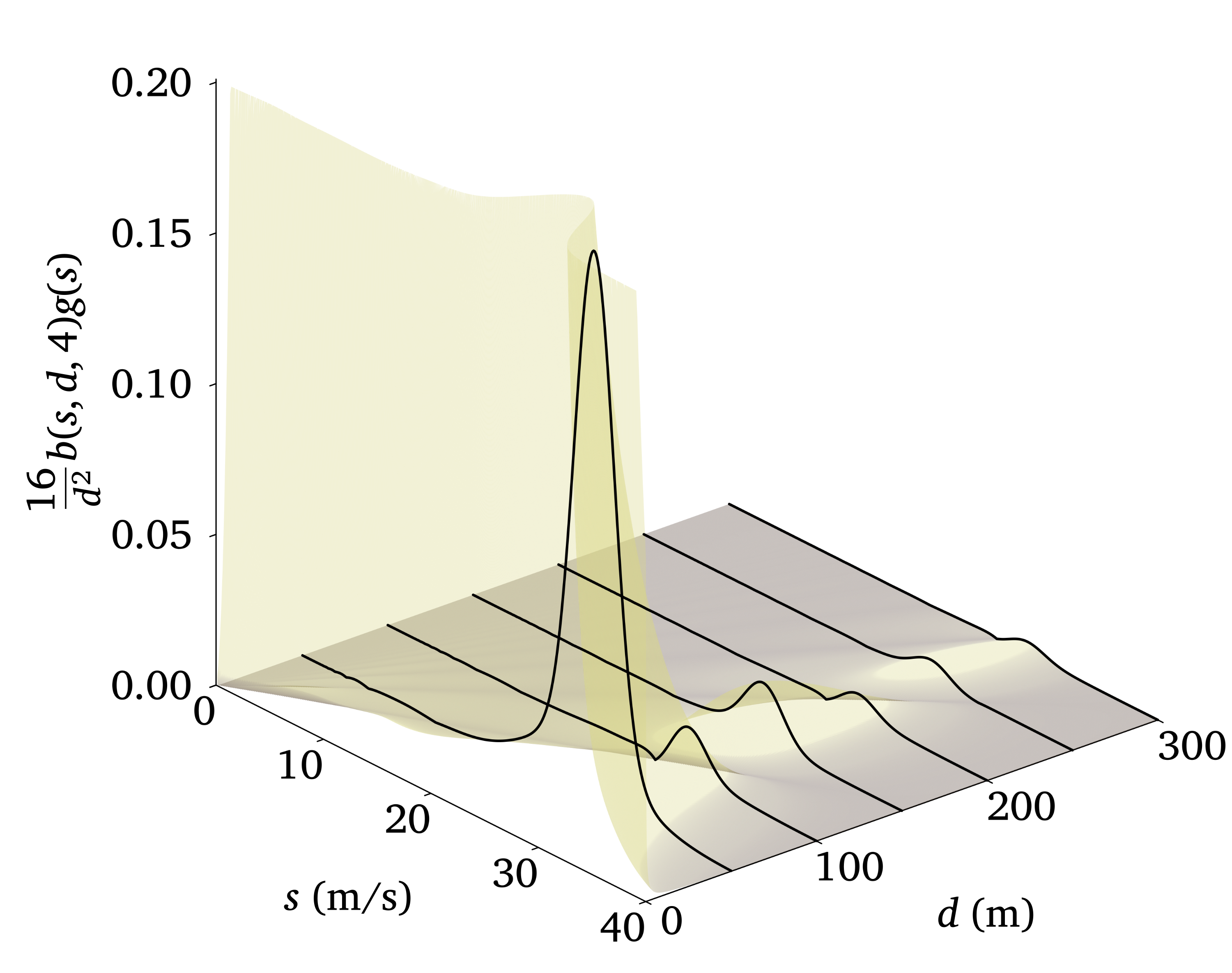}{0.49\textwidth}{(b) $\displaystyle \frac{16}{d^2}b(s, d, 4)g(s)$ \label{fig:figure7b}}}
\caption{Surface plots of $b(s, d, 4)g(s)$ and $\displaystyle \frac{16}{d^2}b(s, d, 4)g(s)$.
\label{fig:figure7}}
\end{figure}

This example provides graphical descriptions of the proof of Corollary 2. Figure \ref{fig:figure7} displays an example: $b(s, d, 4)g(s)$ and $4^2/d^2 \cdot b(s, d, 4)g(s)$ as functions of $s$ and $d$ when $S \sim g(s)$. In Figure \ref{fig:figure7}a, $b(s, d, t)g(s)$ has a periodic pattern along the $d$-axis. Figure \ref{fig:figure7}b is an extension of Figure \ref{fig:figure3}c to the $d$-axis, where $b(s, d, t)g(s)$ is scaled by $t^2/d^2$ to plot Equation \ref{eq:ghvry} when $m = 1$. Because $\mathrm{VMR}[\hat{m}]$ is inversely proportional to $d^2$, a larger $d$ tends to result in a better precision in $\hat{m}$. This is intuitive considering $\mathrm{Var}[\hat{m}]$ arises from the discreteness of the observed number of data points. The ratio of the additional number of data points $K$, a Bernoulli random variable, to the total number of data points $n$ decreases as the cordon captures more data points, owing to a larger $d$.

\begin{figure}[H] 
\gridline{\fig{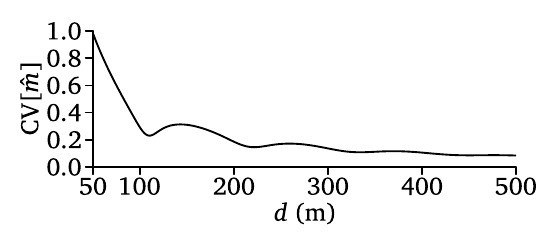}{0.37\textwidth}{(a) CV[$\hat{m}$] as a function of $d$ when $t = 4$ \label{fig:figure8a}}
            \hspace{-2.5cm} 
          \fig{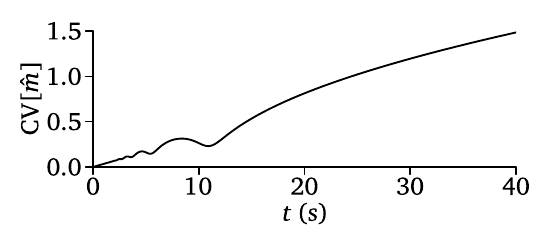}{0.37\textwidth}{(b) CV[$\hat{m}$] as a function of $t$ when $d = 300$ \label{fig:figure8b}}}
\caption{CV[$\hat{m}$] as a function of $d$ and $t$ when the other variables are fixed.
\label{fig:figure8}}
\end{figure}

However, VMR$[\hat{m}]$ or CV$[\hat{m}]$ does not always exhibit a monotonic decrease over $d$. As seen in Figure \ref{fig:figure8}a, the non-monotonicity of $\mathrm{CV}[\hat{m}]$ as a function of $d$ indicates the potential existence of $d$ that locally minimises CV$[\hat{m}]$ when $\max(d)$ exists. When some road geometry dictates $\max(d)$ is 150 m (e.g., a 150-m road segment immediately bounded by intersections beyond which traffic volumes may vary) in the condition of Figure \ref{fig:figure8}a, it would be better to set 110-m $d$ ($CV = 23.048\ \%$) than trying to set 150-m $d$ ($CV = 30.999\ \%$). Figure \ref{fig:figure8}b plots CV$[\hat{m}]$ as a function of $t$ when $d = 300$. CV$[\hat{m}]$ tends to increase as $t$ increases, but this relationship is not always monotonic.

\section{Simulations} \label{subsec:simulation}

We compared numerically simulated distributions of $\hat{m}$ with their theoretical distributions for illustrative purposes\footnote{The simulations are presented solely as a demonstration for the readers. The conclusions of this paper do not rely on the simulation results.}.


\subsection{Method}

In Julia 1.8.5, the number of probe footprints was modelled as a series of particles with independent uniform linear motion along a road segment. We reiterate that $g(s)$ is the true space-mean speed distribution of all probes traversing the cordon and is not necessarily of the free flow speed or target speed that probes were aiming for. In addition, assuming uniform linear motion here is different from assuming that all probes traverse the cordon with uniform linear motion. In this experiment, the emergence of binomial distributions (Equation \ref{eq:yathe2}) was considered trivial. The \textit{Distributions.jl} package \citep{LINetal-2019} was used to generate statistical distributions under the following two scenarios: scenario 1 ($d = 300$ and $t = 4$) and scenario 2 ($d = 40$ and $t = 1$). In each scenario, $m \in \{1, 2, 4, 8\}$ and $S \sim g(s)$ as shown in Figure \ref{fig:figure3}a. We performed one million simulations using Equation \ref{eq:draso} for each combination of scenarios and values of $m$. The simulated distributions were compared to theoretical PDFs.

\subsection{Results} \label{subsubsec:results1}

\begin{deluxetable}{cccccc}[htbp]
    \caption{Descriptive Statistics of $\hat{m}$ in Simulations and Theory}
    \label{tab:table1}
    
    \tablenum{1}
    \tablehead{\colhead{Scenario} & \colhead{$m$} & \colhead{Item} & \colhead{E[$\hat{m}$]} & \colhead{Var[$\hat{m}$]} & \colhead{CV[$\hat{m}$]}} 
    \startdata
    1 & 1 & Simulated & 1.000 & 0.019 & 0.137 \\
     &  & Theoretical & 1 & 0.019 & 0.137 \\
     & 2 & Simulated & 2.000 & 0.037 & 0.097 \\
     &  & Theoretical & 2 & 0.037 & 0.097 \\
     & 3 & Simulated & 4.000 & 0.075 & 0.068 \\
     &  & Theoretical & 4 & 0.075 & 0.068 \\
     & 4 & Simulated & 8.000 & 0.150 & 0.048 \\
     &  & Theoretical & 8 & 0.149 & 0.048 \\
    2 & 1 & Simulated & 1.000 & 0.088 & 0.297 \\
     &  & Theoretical & 1 & 0.088 & 0.297 \\
     & 2 & Simulated & 2.000 & 0.177 & 0.210 \\
     &  & Theoretical & 2 & 0.177 & 0.210 \\
     & 3 & Simulated & 4.000 & 0.353 & 0.148 \\
     &  & Theoretical & 4 & 0.353 & 0.149 \\
     & 4 & Simulated & 7.999 & 0.706 & 0.105 \\
     &  & Theoretical & 8 & 0.706 & 0.105 \\
    \enddata
\end{deluxetable}

\begin{figure}[htbp]
\gridline{\fig{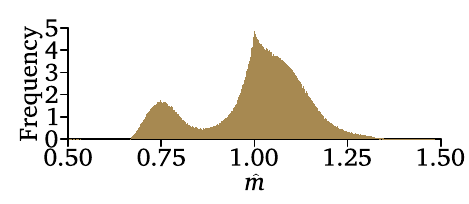}{0.3\textwidth}{\vspace*{-5mm}(a) Histogram of $\hat{m}$ (scenario 1, $m = 1$)\label{fig:figure9a}}\hspace{-5cm} 
          \fig{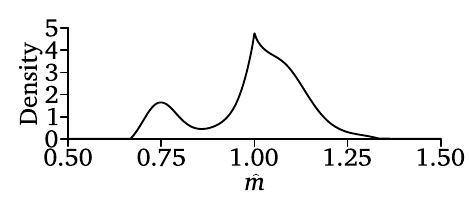}{0.3\textwidth}{\vspace*{-5mm}(b) PDF of $\hat{m}$ (scenario 1, $m = 1$)\label{fig:figure9b}}}\vspace*{-7mm}
\gridline{\fig{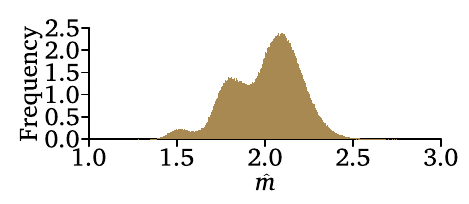}{0.3\textwidth}{\vspace*{-5mm}(c) Histogram of $\hat{m}$ (scenario 1, $m = 2$)\label{fig:figure9c}}\hspace{-5cm}
          \fig{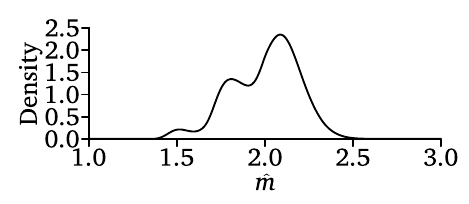}{0.3\textwidth}{\vspace*{-5mm}(d) PDF of $\hat{m}$ (scenario 1, $m = 2$)\label{fig:figure9d}}}\vspace*{-7mm}
\gridline{\fig{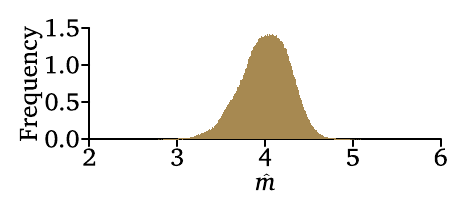}{0.3\textwidth}{\vspace*{-5mm}(e) Histogram of $\hat{m}$ (scenario 1, $m = 4$)\label{fig:figure9e}}\hspace{-5cm}
          \fig{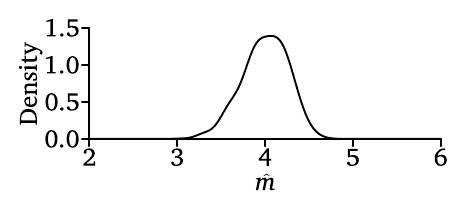}{0.3\textwidth}{\vspace*{-5mm}(f) PDF of $\hat{m}$ (scenario 1, $m = 4$)\label{fig:figure9f}}}\vspace*{-7mm}
\gridline{\fig{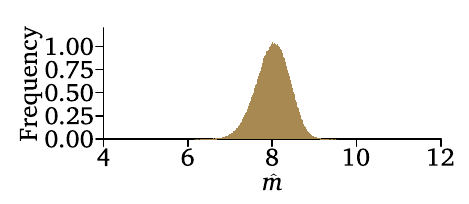}{0.3\textwidth}{\vspace*{-5mm}(g) Histogram of $\hat{m}$ (scenario 1, $m = 8$)\label{fig:figure9g}}\hspace{-5cm}
          \fig{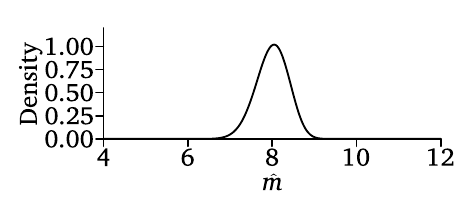}{0.3\textwidth}{\vspace*{-5mm}(h) PDF of $\hat{m}$ (scenario 1, $m = 8$)\label{fig:figure9h}}}\vspace*{-7mm}

\gridline{\fig{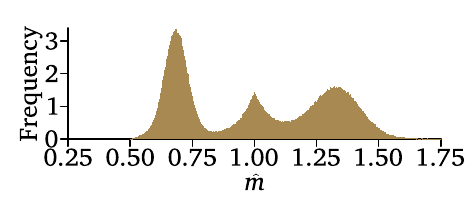}{0.3\textwidth}{\vspace*{-5mm}(i) Histogram of $\hat{m}$ (scenario 2, $m = 1$)\label{fig:figure9i}}\hspace{-5cm}
          \fig{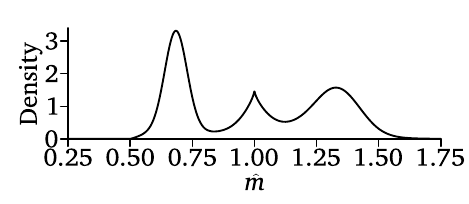}{0.3\textwidth}{\vspace*{-5mm}(j) PDF of $\hat{m}$ (scenario 2, $m = 1$)\label{fig:figure9j}}}\vspace*{-7mm}
\gridline{\fig{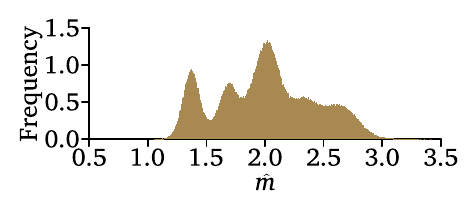}{0.3\textwidth}{\vspace*{-5mm}(k) Histogram of $\hat{m}$ (scenario 2, $m = 2$)\label{fig:figure9k}}\hspace{-5cm}
          \fig{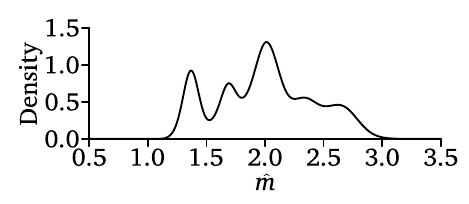}{0.3\textwidth}{\vspace*{-5mm}(l) PDF of $\hat{m}$ (scenario 2, $m = 2$)\label{fig:figure9l}}}\vspace*{-7mm}
\gridline{\fig{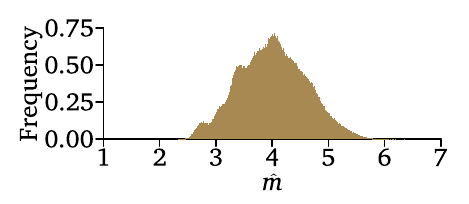}{0.3\textwidth}{\vspace*{-5mm}(m) Histogram of $\hat{m}$ (scenario 2, $m = 4$)\label{fig:figure9m}}\hspace{-5cm}
          \fig{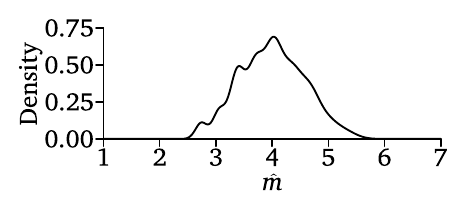}{0.3\textwidth}{\vspace*{-5mm}(n) PDF of $\hat{m}$ (scenario 2, $m = 4$)\label{fig:figure9n}}}\vspace*{-7mm}
\gridline{\fig{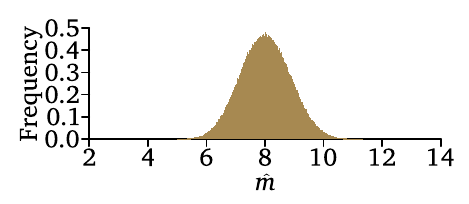}{0.3\textwidth}{\vspace*{-5mm}(o) Histogram of $\hat{m}$ (scenario 2, $m = 8$)\label{fig:figure9o}}\hspace{-5cm}
          \fig{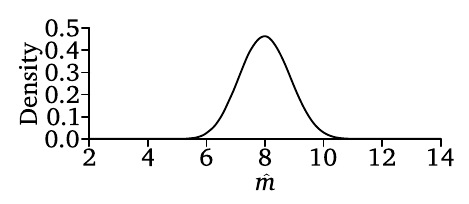}{0.3\textwidth}{\vspace*{-5mm}(p) PDF of $\hat{m}$ (scenario 2, $m = 8$)\label{fig:figure9p}}}\vspace*{-0mm}
\caption{Histograms of simulated $\hat{m}$ and theoretical PDFs of $\hat{m}$.
\label{fig:figure9}}
\end{figure}

Table \ref{tab:table1} exhibits the descriptive statistics of simulations and theoretical values, while Figure \ref{fig:figure9} shows the histograms of simulated $\hat{m}$ and theoretical PDFs of $\hat{m}$ calculated by Equation \ref{eq:hnmnmi}. The simulation results showed a good match in descriptive statistics between simulated and theoretical values.

As shown in Figure \ref{fig:figure9}, $\hat{m}$ distributes around $m$, but the PDFs are not necessarily line-symmetric around $\hat{m} = m$.
The PDFs approached normal distributions as $m$ increased.


\section{Implication of the Model} \label{subsec:implicationofthemodel}

This paper presented the exact distribution of estimated probe traffic volume $\hat{m}$ based on the point probe location data recorded at a fixed interval. The final section discusses the model’s implications regarding theory, applications, and opportunities.

\subsection{Model Characteristics} \label{subsec:modelcharacteristics}

Practitioners can use $\hat{m}$ as an unbiased estimator of probe traffic volumes in any timeframe. The more probes are present, the more closely the distribution of $\hat{m}$ can be approximated by a normal distribution. Equation \ref{eq:draso} alone can give $\hat{m}$ as an estimate of $m$, but $\mathrm{Var}[\hat{m}]$ guides how the analyst should set the cordon. The estimation imprecision, measured as CV[$\hat{m}$], is inversely proportional to the square root of the actual probe volume $m$, roughly proportional to recording interval $t$, and roughly inversely proportional to cordon length $d$ (Equation \ref{eq:hokeg}). In other words, the higher the probe volume, the more precise the volume estimates are likely to be, while the degree of marginal improvement decreases as the traffic volume increases. A lower probe speed also tends to yield better precision when other conditions are held constant.

The relationship between $d$ and CV[$\hat{m}$] is not always monotonic. Depending on the recording interval and speed distribution, there is a local optimal cordon length $d$ that maximises the precision of $\hat{m}$ estimation (i.e., minimises $\mathrm{CV}[\hat{m}]$) (Figure \ref{fig:figure8}a).
Although the authors are unaware of the exact data processing methods used in proprietary traffic volume estimation software, the estimation precision is likely to improve by setting an optimal cordon length $d$ in these products if the software inherently relies on probe point data with speed information.
It should be noted that the sensitivity analysis, as discussed in Example 4, does not hold when $g(s)$ drastically changes with $d$ (e.g., a segment with high speed shear). In practice, the speed distribution $g(s)$ could change along with $d$; thus, the theoretical optimal cordon length $d$ should be seen a suggestion rather than a perfect means of optimisation. Therefore, it is a reasonable strategy to set the longest possible $d$ that fits the road segment that carries a single probe traffic volume when an analyst does not have complete information about the probe data recording interval $t$ or the speed distribution $g(s)$.

If one desires to use $\hat{m}$ as a means of traffic volume estimation, calibration of $\hat{m}$ is required to convert these values into traffic volume estimates. Because probes are unlikely to be distributed homogeneously among road users, this procedure ultimately determines traffic volume estimation accuracy. During this process, modellers can use $1/\mathrm{Var}[\hat{m}]$ as a weight of each $\hat{m}$ to maximise traffic volume estimation accuracy \citep{AITKEN-1935}.

The proposed method can be applied to probe point datasets, provided they can be separated by homogeneous $t$. When a data integrator has probe point data from mixed sources with various $t$, the proposed method is applicable only upstream of the data processing; namely, before mixing probe data from multiple sources. Once $\hat{m}$ is obtained for each $t$, the values of $\hat{m}$ can be further integrated using $1/\mathrm{Var}[\hat{m}]$ as weights.

\subsubsection{Limitations} \label{subsec:limitations}

Practitioners should be aware of limitations when applying the proposed method to probe point location data. First, spatial characteristics should be considered when drawing virtual cordons. For example, a modeller must pay attention to grade-separated facilities, tunnels, crosswalks, sidewalks, and cell phone location data from flying objects. Sometimes, probe data need to be coded to avoid capturing location data from unintended road users, as we truncated the high speed in our example.

In the absence of measurement errors, $1/\mathrm{Var}[\hat{m}]$ gives the theoretical upper bound on the precision of probe traffic volume estimation. With real traffic, $\mathrm{Var}[\hat{m}]$ can become larger than the theoretical one because GNSSs are not free from systematic and random measurement errors \citep{MARKOVICetal-2019}. The degree of deterioration in estimation precision due to measurement errors will depend on $d$, $t$, and the accuracy of GNSS. Although centimetre-level positioning is available with some GNSSs \citep{CHOYetal-2015}, GNSS argumentation is associated with horizontal errors varying up to 3--15 m \citep{MERRYandBETTINGER-2019, ZANDBERGENandBARBEAU-2011}. As a result, speed measurement is also associated with some errors \citep{GUIDOetal-2014}. Generally speaking, the longer $d$ is, the more the random error is expected to cancel out. For this reason, it would be reasonable to set a long $d$ when it is possible. Because speed distribution plays a crucial role in estimating traffic volumes in the proposed method, it is essential to make an effort to reduce speed bias \citep{AHSANIetal-2019} in the data acquisition process. For example, the speed of a stationary probe could be incorrectly recorded as a small positive number instead of zero due to GNSS measurement errors. When this happens, $\hat{m}$ calculated by Equation \ref{eq:draso} becomes larger than it should be. While this is not a theoretical flaw, some preprocessing, such as considering speeds below a certain threshold zero, may be necessary in practical settings.

In traffic volume estimation, another limitation of the model is that the PDF formulation (Equation \ref{eq:hnmnmi}) of $\hat{m}$ includes the true probe volume $m$ itself. Although this does not prevent the computation of $\hat{m}$ (Equation \ref{eq:draso}) or VMR[$\hat{m}$] (Equation \ref{eq:ghvry}), this recursion is sometimes not ideal, because the probe volume is usually estimated when the probe volume $m$ is unknown. In this context, this study is theoretical and may not serve as a silver bullet for all issues readers expect to be solved.

\subsection{Applications} \label{subsec:applications}

The proposed method can contribute to various aspects of traffic volume estimation. First, it allows agencies to use marginal point probe data without pseudonyms or granular timestamps. For example, they can enhance the quality of traffic volume estimation by utilising sparsely recorded probe data, which would have been ignored without our method. Depending on how much marginal probe point data are available compared with the line data already available, probe location data without pseudonyms can be a sleeping lion.

The theoretical aspect of the distribution of estimated probe traffic volume based on point data is meaningful not only for deepening our understanding of the ever-increasing probe location data but also for unfolding the mechanisms that tend to be obscured in machine learning. It is preferable for models to have some degree of explainability rather than accepting machine learning models without thorough understanding, especially when public funds are involved \citep{ROLL-2023}. As reported by \cite{TURNER-2021}, the explainability and evaluation of big data quality and valuation, however, have been of concern among transportation professionals, as machine learning models can quickly become black boxes. The theoretical distribution of $\hat{m}$ is valuable in this context because it partially explains, even with some measurement errors, the mechanisms behind traffic estimation models developed by directly applying machine learning models to probe point data without estimating $m$. In certain situations, such as road segments with low speed shear, this knowledge can enhance the traffic volume estimation models, as illustrated in Figure \ref{fig:figure8}. The proposed method enables modellers to efficiently incorporate low probe volumes into their traffic volume estimation models. The theoretical PDF of the estimated probe traffic volume allows modellers and analysts to perform interval estimation on $m$. Depending on the calibration model, probe traffic volume estimates with confidence intervals (CIs) can also be used to improve the calibration accuracy against known traffic volumes. Also, the proposed model hints that the distribution of $\hat{m}$ can be used to estimate the valuation of probe point data. From Equation \ref{eq:ghvry}, it may, for example, be reasonable to value point probe data as approximately inversely proportional to $t^2$.

Furthermore, the model predicts “economies of scale”, encompassing probe data valuation. A higher recording frequency ($\because$ Equation \ref{eq:hokeg}) and homogeneity make the traffic volume estimation more precise and accurate, respectively. As a result, probe location data with high recording frequency and homogeneity are more valuable for traffic volume estimation. Thus, agencies could perform cost-benefit analyses based on the specific goals they want to achieve.

Another economy of scale arises from the synergistic effect of acquiring traffic counts at fixed locations. Probe traffic volumes can be used to estimate traffic volumes at many locations. This fact does not diminish the importance of fixed-location traffic counts, because it is impossible to calibrate the values against traffic volumes without ground truths. A higher density of reliable traffic count data from conventional devices can enhance the proposed method by providing additional calibration data. Therefore, governments investing in continuous traffic monitoring infrastructure can expect an even larger return on investment (ROI) than they expect.

\subsection{Opportunities} \label{subsec:opportunities}

The proposed technique can positively impact society, as transportation systems are woven into daily human activities. On a global scale, traffic volume estimations based on probe point data can positively impact agencies and nations with limited financial and human resources \citep{LORDetal-2003, YANNISetal-2014}. The method will be particularly useful for low-volume rural roads, where traditional traffic counting tools may not be cost-efficient \citep{DAS-2021}. Because remote highways tend to have long uninterrupted segments \citep{LORDetal-2011}, drawing long virtual cordons can help transportation professionals estimate probe traffic volumes with great precision. Traffic volume information along rural highways can be used to develop safety performance functions (SPFs) more thoroughly and continuously than ever before \citep{TSAPAKISetal-2021b}.

Because traffic volume estimation using probe data is in its infancy, there are many research opportunities in this field. From a practical standpoint, future research related to traffic volume estimation from probe point data would include the formulation of an error term for speed measurement in the distribution of $\hat{m}$, the development of universal indices to describe the homogeneity of probe data, a framework for evaluating data transferability, cost-benefit analyses of probe location data, and real-time crash hotspot identification.

Our model paves the way for unleashing probe point data for social good. In the 1940s, \cite{GREENSHIELDS-1947} analysed traffic using a series of aerial photographs taken at fixed intervals. Decades later, we have the opportunity to improve the quality of transportation through “snapshots” of probes recorded at fixed intervals with unprecedented scalability. Inter-organisational collaborations, including cooperation between the public and private sectors, will be crucial for bringing the technology to life.

\section*{Glossary}
\begin{itemize}
    \item \textit{Line data} -- A series of chronologically connected point data.
    \item \textit{Point data} -- Data that contain information to identify a point location on a surface.
    \item \textit{Probe} -- A device that records its position as point data in the Earth's spatial reference system (e.g., geographic coordinates). Probes (e.g., smartphones) are not limited to vehicles.
    \item \textit{Probe traffic volume} -- The number of probes traversing a cross-section.
\end{itemize}

\section*{Acknowledgements}
The first author would like to express gratitude to Dr. Daniel Romero at the University of Agder for his valuable advice in the field of statistics.

\section*{Funding Source Declaration}
This research was funded in part by the A.P. and Florence Wiley Faculty Fellow provided by the College of Engineering at Texas A\&M University.

\bibliography{sample631}{}
\bibliographystyle{chicago}

\end{document}